\documentclass[useAMS,usenatbib]{mn2e}
\usepackage{lscape}
\input psfig.sty
\input epsf.sty
\usepackage{graphicx}
\usepackage{epsfig}

\title[The Spatial Distribution of Star Formation in the Solar Neighbourhood: Do all stars form in clusters?]{The Spatial Distribution of Star Formation in the Solar Neighbourhood: Do all stars form in dense clusters?} 
\author[Eli Bressert et al.]{E. Bressert$^{1,2}$\thanks{E-mail: eli@astro.ex.ac.uk (EB)}, N. Bastian$^{1,3}$, R. Gutermuth$^{4}$, S.T. Megeath$^{5}$, 
L. Allen$^{6}$, \newauthor Neal J. Evans II$^{7}$, L.M. Rebull$^{8}$, J. Hatchell$^{1}$, D. Johnstone$^{9,10}$, T.L. Bourke$^{2}$,\newauthor 
L.A. Cieza$^{12}$, P.M. Harvey$^{7}$, B. Merin$^{13}$, T.P. Ray$^{14}$, \& N.F.H. Tothill$^{1,15}$\\
$^{1}$School of Physics, University of Exeter, Stocker Road, Exeter EX4 4QL, UK\\
$^{2}$Harvard Smithsonian Center for Astrophysics, 60 Garden St., Cambridge MA 02138, USA\\
$^{3}$Institute of Astronomy, University of Cambridge, Madingley Road, Cambridge CB3 0HA, UK\\
$^{4}$University of Massachusetts, Smith College, Northampton, MA 01063, USA\\
$^{5}$Department of Physics \& Astronomy, MS-113, University of Toledo, 2801 W. Bancroft St., Toledo, OH 43606, USA\\
$^{6}$National Optical Astronomy Observatory, 950 North Cherry Avenue, Tucson, AZ 85719, USA\\
$^{7}$Department of Astronomy, University of Texas at Austin, 1 University Station C1400, Austin, TX, USA\\
$^{8}$Spitzer Science Center/Caltech, M/S 220-6, 1200 East California Boulevard, Pasadena, CA 91125, USA\\
$^{9}$National Research Council Canada, Herzberg Institute of Astrophysics, 5071 West Saanich Road, Victoria, BC V9E 2E7, Canada\\
$^{10}$Department of Physics \& Astronomy, University of Victoria, Victoria, BC, V8P 1A1, Canada\\
$^{12}$Institute for Astronomy, University of Hawaii at Manoa, Honolulu, HI 96822, USA\\
$^{13}$Herschel Science Center, European Space Agency (ESA), P.O. Box 78, 28691 Villanueva de la Ca\~{n}ada (Madrid), Spain\\
$^{14}$School of Cosmic Physics, Dublin Institute for Advanced Studies, Republic of Ireland\\ 
$^{15}$School of Physics, University of New South Wales, Sydney, NSW 2052, Australia}

\begin{document}

\maketitle \label{firstpage}
\begin{abstract}
    We present a global study of low mass, young stellar object (YSO) surface densities ($\Sigma$) in nearby ($< 500$~pc) star forming regions based on a comprehensive collection of {\it Spitzer Space Telescope} surveys.  We show that the distribution of YSO surface densities in the solar neighbourhood is a smooth distribution, being adequately described by a lognormal function from a few to $10^3$ YSOs per $\textrm{pc}^{2}$, with a peak at $\sim22$~stars $\textrm{pc}^{-2}$ and a dispersion of $\sigma_{{\rm log}_{10} \Sigma}$ $\sim$ 0.85.  We do not find evidence for multiple discrete modes of star-formation (e.g.~clustered and distributed).  Comparing the observed surface density distribution to previously reported surface density threshold definitions of clusters, we find that the fraction of stars in clusters is crucially dependent on the adopted definitions, ranging from $40$ to $90$\%. However, we find that only a low fraction ($< 26$\%) of stars are formed in dense environments where their formation/evolution (along with their circumstellar disks and/or planets) may be affected by the close proximity of their low-mass neighbours. 

\end{abstract}

\section{Introduction} \label{sec:introduction}

It is often stated that most if not all stars form in stellar clusters. This view is based largely on near-infrared (NIR) studies of star-forming (SF) regions within several hundred parsecs of the Sun \citep{lada03,porras03}. However, adding high-resolution mid-infrared (MIR) data to the NIR makes YSO identification more robust and less likely to be contaminated by field stars, which leads to better tracing of YSO surface densities. This means that with the NIR alone, there were large uncertainties in the number of stars at low values of YSO surface densities ($\Sigma_{\rm YSO}$) \citep{carpenter00}.

The spatial distribution of forming stars, i.e. do they form in clusters, is important for two main reasons. The first is that dense environments can affect the evolution of the young stars as well as alter their disk and planet formation/evolution \citep{allen07}.  The second is to locate the progenitor population of open clusters and to determine why such a low fraction of the Galactic stellar population is observed in clusters. Are there multiple discrete modes, such as clustered and distributed, in the star-formation process that manifest themselves as peaks in a surface density distribution (e.g. \citealt{strom93,carpenter00,weidner04,wang09})?

With the launch of the {\it Spitzer Space Telescope} \citep{werner04} we are now able to differentiate YSOs and contaminating sources based on colour information and hence can study the distribution of YSOs independently of the surface densities. Large field-of-view (FoV) {\it Spitzer} observations of SF regions \citep{allen07,evans09} found that YSOs extend well beyond the densest groups in their environment and continue throughout. We combine several {\it Spitzer} surveys that cover nearly all the SF regions within 500~pc of the Sun. A list of the regions and their properties is given in Table \ref{tab:objects}. Note that with only the local SF environments being considered, we are not sampling massive star forming regions that are found beyond 500~pc.

Using the comprehensive collection of $\Sigma_{\rm YSO}$ we investigate what fraction of YSOs are found in dense clusters. We define dense clusters as regions where YSOs are affected by their neighbours in sufficiently short timescales of $<10^5$ yr, such that its surface densities exceed $\sim200$ $\textrm{YSO}$ pc$^{-2}$ (see \citealt{gutermuth05}). We also review what surface densities are required to identify ``clusters" according to definitions provided by \cite{carpenter00,lada03,allen07,jorgensen08,gutermuth09} in \S~\ref{sec:identification}. In this paper we will investigate 1$)$ whether there is evidence for multi-modality in the surface densities of YSOs, 2$)$ what fraction of stars form in dense clusters in the local neighbourhood and 3$)$ how relevant the various cluster definitions are.

\begin{table} 
    \begin{tabular}
        {lccr} Name&YSO Number&Distance&Reference\\
        &  & \textit{pc} & \\
        \hline Auriga&138(172)&300&1\\
        Cepheus I & 34(46)&280&1\\
        Cepheus III&44(52)&280&1\\
        Cepheus V&19(19)&280&1\\
        Chameleon I&67(93)&200&1\\
        Corona Australis&27(45)&130&1\\
        Lupus III&43(79)&150&2\\
        Ophiuchus&199(297)&125&2\\
        Orion$^{\star}$&2696(3352)&414&3\\
        Perseus&280(387)&250&2\\
        Serpens&179(262)&415&2\\
        Taurus&131(249)&137&4\\
        \hline 
    \end{tabular}
    \caption{The {\it Spitzer} surveys used in the present work includes 12 star forming regions with 3857 YSOs. The numbers in brackets refer to the total number of sources in the catalogues for each region, while the number before the brackets is the number used in the present analysis.  The difference is due to the application of the absolute magnitude cuts as well as the elimination of class~III YSOs from the sample. The sources for these SF regions are the 1$)$ GB survey, 2$)$ c2d survey, 3$)$ Orion survey and 4$)$ Taurus survey.}
$^{\star}${\small ONC is excluded, see \S~\ref{sec:observations}.}
\label{tab:objects}
\end{table}

\section{Observations \& Data} \label{sec:observations}

Multiple {\it Spitzer} surveys were used to generate a comprehensive and statistically significant dataset to investigate the spatial surface density properties of forming stars in the solar neighbourhood. The surveys are the Gould's Belt (GB) survey (Allen et al. in prep.), Orion survey (Megeath et al. in prep.), Cores to Disks (c2d) survey \citep{evans03}, and the Taurus survey \citep{rebull10}. The GB and Orion catalogs have not been publicly released yet. We have more than 7000 YSO detections in the combined catalogs at distances between 100 to 500~pc.

{\it Spitzer} data are necessary for this study as low $\Sigma_{\rm YSO}$ can be differentiated from field star populations, unlike NIR observations where field star contamination can be problematic. The YSO population that we have collected represents a global view of the low-mass star-forming region in the local neighbourhood from low to high surface densities. These {\it Spitzer} surveys combined represent the most complete census of star formation within 500~pc of the Sun available to date.

In order to homogenise the data from the surveys we accounted for distance effects on photometry, namely we limit the absolute magnitude range used for individual sources to that of the faintest YSO detectable in the furthest SF region and the brightest in the nearest SF region. The absolute magnitude limit used for the 500~pc data collection is 0 $\leq$ M$_{3.6\mu m}$ $\leq$ 5.91, based on Orion at a distance of 414~pc \citep{menten07,mayne08} for the faint sources and Ophiuchus at 125~pc for the bright sources. This reduces the number of YSOs we can use, but it mitigates detection biases introduced for SF regions at different distances.

The GB and c2d surveys classify YSOs using spectral indices \citep{lada87,greene94}. The Taurus and Orion YSOs are classified by using colour-colour diagrams \citep{allen04, megeath04, gutermuth05, gutermuth09}. What fraction of the YSOs are diskless, generally classified as Class III, and hence not identifiable in the IR? Based on \cite{hernandez07} we assume that 65\% of the YSOs have disks. We corrected the stellar surface densities of the data for the missing fraction of 35\%.

Orion, which offers the largest range of stellar surface densities and hosts the most massive stars of the SF regions considered in this paper, had to be treated separately from the other surveys. The ONC, in particular the Trapezium region, has two {\it Spitzer} based issues: Stellar surface densities that exceed {\it Spitzer's} spatial resolution and the extremely bright nebulosity that diminishes effective sensitivity considerably. The bright nebulosity introduces errors for YSO identification since the PAH emission outshines lower mass YSOs and introduces large errors in the photometry. To compensate for the complex incompleteness we removed all YSOs centred on $\Theta1$ Orionis within a radius of 0.56~pc (4.7'). To correct for missing YSOs from the removed region, we estimated that the mass removed was $\sim25$\% of the total Orion complex \citep{getman05}. Excluding the ONC from our analysis does not significantly change the presented cumulative distribution of surface densities presented in this paper. If we were able to observe all the members in the ONC based on the $\sim25$\% of mass we estimated to be missing, the average $\Sigma_{\tiny \textrm{ONC}}$ $\leq 1000$~YSOs/$\textrm{pc}^{2}$. This surface density regime goes beyond the scope of values we are presently considering. Hence we are not sensitive to the extreme high $\Sigma$ tail end of the ONC distribution.

{\it Spitzer} is not completely free of contamination when identifying YSOs, i.e. AGBs/Be stars \citep{cieza10, robitaille08} and galaxies \citep{gutermuth08,evans09}. \cite{oliviera09} found that $\sim 25\%$ of the identified YSOs in the c2d Serpens catalog are AGBs, which is likely an isolated worst case scenario as Serpens is the field closest to the Galactic plane in our compilation of SF regions. Two of the twenty contaminants \cite{oliviera09} identified are Class IIs and the rest of the contaminants are Class IIIs. We only consider Class I/II objects, where the AGB contamination is $< 10\%$, and remove all Class IIIs. The flat spectrum sources are grouped with Class I objects. Between the methods used to identify YSOs in the c2d, GB, Taurus and Orion data, which are the c2d \citep{evans09} and Gutermuth et al. (2008, 2009) methods, the selection discrepancy is $\leq 5\%$ \citep{rebull10}. By selection discrepancy we mean the agreement that an object is or is not a YSO (Class I/II).

Extra-galactic background contamination for YSO MIR identification is well studied. For the c2d and GB catalogs, which use the same data-reduction pipeline, \cite{evans09} found that background galaxies contaminate $\leq 5\%$ of the YSOs. Similarly, YSOs identified via the \cite{gutermuth09} method for Orion is $<1\%$. For Taurus the expected contamination rate is $\leq 5\%$ \citep{rebull10}.

\begin{figure}
    \includegraphics[width=8cm]{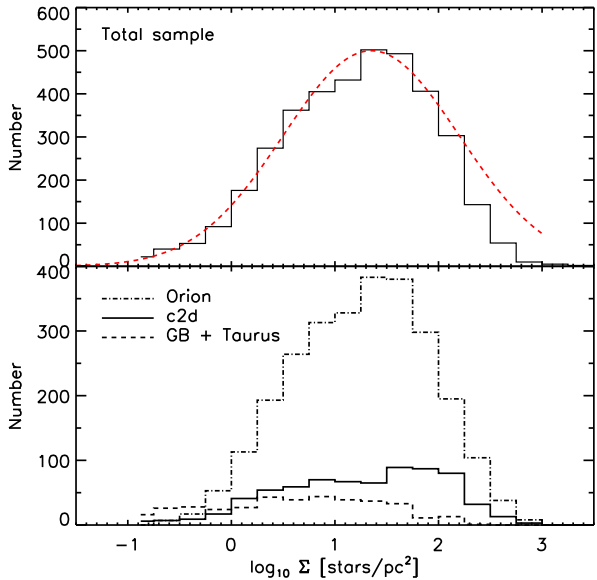} \caption{{\bf Top panel:} The surface density distribution of the total sample of YSOs in the solar neighbourhood used in this work (black).  A lognormal function with a peak at $\sim22$~YSOs/\textrm pc$^2$ and a dispersion $\sigma_{{\rm log}_{10} \Sigma} = 0.85$ is shown as a dashed (red) line. {\bf Bottom panel:} The same as the top panel but now broken into the three respective surveys.  Note that Orion dominates the number statistics.}
    \label{fig:histogram} 
\end{figure}

\begin{figure}
    \includegraphics[width=8cm]{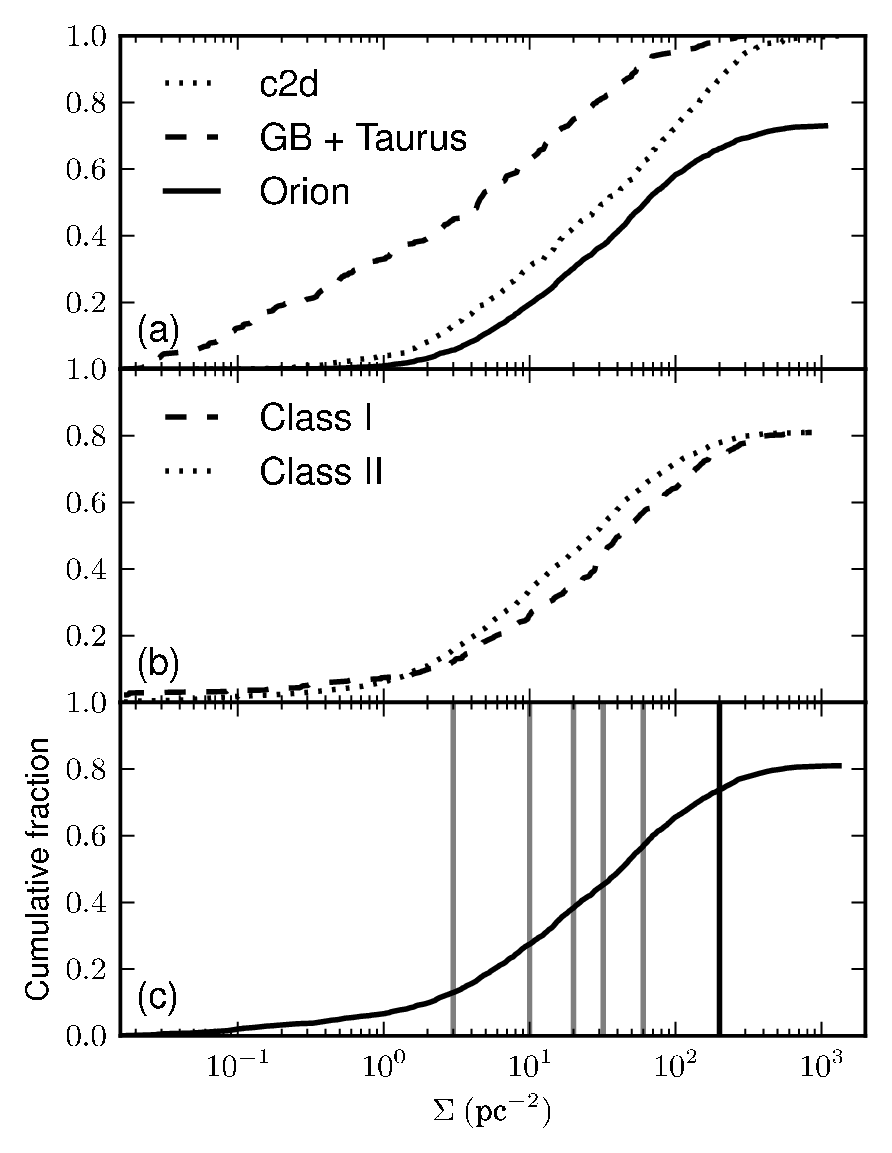} \caption{{\bf(a)} The cumulative fraction of surface densities for the GB+Taurus, c2d, and Orion surveys. Each SF region included in the distributions has N(YSOs) $\geq 10$ and a sufficient field-of-view to properly calculate stellar surface densities. The Orion survey stops at 73\% for the cumulative fraction since the ONC is excluded. We adopt a 65\% disk fraction for all of the SF regions.  We normalised each curve by the number of YSOs in each survey. {\bf(b)} With the GB+Taurus, c2d, and Orion surveys combined we see Class I \& II distributions having similar profiles with a small offset in density, showing that we are likely seeing the primordial distribution of the YSOs. {\bf(c)} With all of the {\it Spitzer} surveys combined we compare several cluster definitions. The vertical grey lines from left to right are Lada \& Lada (2003), Megeath et al. (in prep.), J{\o}rgensen et al. (2008), Carpenter (2000), and Gutermuth et al. (2009) stellar density requirements for clusters. These values correspond to 3, 10, 20, 32, and 60 YSOs $\textrm{pc}^{-2}$ and intersect the corrected cumulative distribution profile, implying that 87\%, 73\%, 62\%, 55\%, and 43\% of stars form in clusters, respectively. The percentages correlate to what fraction of stars form in ``clusters'' based on the various definitions. The black vertical line is for a dense cluster where $\Sigma \geq$ 200 YSOs$/\textrm{pc}^{2}$. The fraction of YSOs in a dense cluster is $<26$\%.} \label{fig:4box}
\end{figure}

\section{$\Sigma_{\rm YSO}$ distributions} \label{sec:biases}
Our primary tool for analysing the surface densities is computing the local observed surface density of YSOs centred on each YSO's position, where $\Sigma_{\rm YSO} = (\textrm{N}-1)/(\pi D_{N}^2)$ and N is the Nth nearest neighbour, and $D_{N}$ is the projected distance to that neighbour (see \citealt{casertano85}). Throughout this work we will adopt $\textrm{N}=7$, although we note that all results have been tested for $\textrm{N}=4-22$ and no significant differences were found.

Figure~\ref{fig:histogram} shows the surface density distribution of all YSOs in our sample, corrected for the diskless fraction.  Additionally, we show a lognormal fit to the data as a dashed red line (see \S~\ref{sec:results}). The over-prediction of the lognormal at high $\Sigma_{\rm YSO}$ compared to the observations is most likely due to the exclusion of the ONC and surrounding area (see \S~\ref{sec:observations}). The bottom panel of Fig.~\ref{fig:histogram} shows the surface density distribution for each of the three surveys separately.

In order to see the fraction of YSOs above a given $\Sigma$ threshold, we show the combined $\Sigma_{\rm YSO}$ distribution (shown as a cumulative fraction normalised to the number in each combined survey) for the three surveys used in this study in Fig.~\ref{fig:4box}a. Note that the GB/Taurus distribution lies to the left of the c2d survey. This is simply due to the GB/Taurus focussing on lower density regions than c2d. The cumulative distribution for the Orion survey only reaches 0.73 in Fig.~\ref{fig:4box}a and 0.81 in Fig.~\ref{fig:4box}c, where all the surveys have been combined, due to the exclusion of the ONC. In Fig.~\ref{fig:4box}c we show the cumulative distribution of all YSOs included in our survey, while in Fig.~\ref{fig:4box}b we split the survey into class I and class II objects.

\section{Results} \label{sec:results} 

It has been long assumed that two distinct modes of star formation exist for YSOs, `clustered' and `distributed' (e.g. \citealt{gomez93,carpenter00,lada03}), but the notion has been questioned after \emph{Spitzer} results hinted otherwise \citep{allen07}. If there are indeed two modes, then we would expect to see a bi- or multi-modal profile in cumulative surface density distribution plots such as Figs. ~\ref{fig:histogram}, ~\ref{fig:4box}a, \& ~\ref{fig:4box}c. Instead we see smooth and featureless distributions from the low to high stellar surface densities for the c2d, GB, Taurus, and Orion surveys. We find that the $\Sigma_{\rm YSO}$ distribution of low-mass stars in the solar neighbourhood can be well described by a lognormal function, as seen in Fig.~\ref{fig:histogram}, with a peak at $\sim22$~YSOs/\textrm pc$^2$ and a dispersion $\sigma_{{\rm log}_{10} \Sigma} = 0.85$.

The spatial distribution of the YSOs in these SF regions is expected to be close to primordial since their YSOs, in particular Class Is and Class IIs, are $\le 2$ Myr old \citep{haisch01,hernandez07}. In order to place stricter constraints on this, we now split the complete sample into Class I and II objects, which can be roughly attributed to an age sequence. The cumulative $\Sigma$ distributions of Class I and II YSOs are shown in Fig.~\ref{fig:4box}b. We see that the two distributions have similar smooth density spectra, however they are slightly offset. The $\Sigma$ of the Class I/II objects are calculated by finding a YSO's Nth nearest YSO. Once this is done for the YSOs we separate the Class I/II objects. $\Sigma$ is calculated this way since Class Is and Class IIs  are not always spatially distinct from one another \citep{gutermuth09}. Class IIs are known to be slightly more dispersed than Class Is in high density regions \citep{gutermuth09} reflecting early dynamical evolution. However, the similar distribution between these classes leads us to conclude that the distribution of observed $\Sigma$ is mainly primordial in nature.

\section{Cluster Identification} \label{sec:identification} 
The definitions of what defines a cluster vary widely as we have limited knowledge about YSO membership other than their projected two dimensional spatial distributions. Some definitions have a physical motivation (e.g. \citealt{lada03}) while the others are generally empirically-derived from the data being considered \citep{allen07}. When applied to a uniform dataset like ours, differing choices of a surface density threshold returned different ``clustered fractions'', as summarised below.\\
\\
\textbf{\cite{carpenter00}:} Clusters in SF regions are identified by using stellar density maps in the $K_{s}$ band. The density maps are field-star background-subtracted (galactic coordinate dependent) based on semi-empirical models.  Clusters are identified as 2$\sigma$ over-densities and defined as regions with $6\sigma$ over-densities (with total number of members taken as the number of sources above the $4\sigma$ threshold) with respect to the local background. Carpenter's cluster $\Sigma_{\rm YSO}$ ranged from 20 to 67 YSOs $\textrm{pc}^{-2}$ with a median of 32 YSOs $\textrm{pc}^{-2}$. Considering the median, 55\% of the YSOs are contained in clusters.\\
\\
\textbf{\cite{lada03}:} A physically related group of stars, called an {\it embedded cluster}, that is 1$)$ partially or fully enshrouded in interstellar gas and dust, 2$)$ has $\geq$ 35 YSOs and 3$)$ a stellar-mass volume density of 1.0 $ M_{\sun}$ $\textrm{pc}^{-3}$ or greater such that its evaporation time exceeds $10^8$ years. In surface density, rather than volume density, the number of YSOs $\textrm{pc}^{-2}$ necessary for ``cluster'' is $\sim 3$ (see \citealt{jorgensen08}). The authors estimated that 80-90\% of the YSOs are in embedded clusters, which is found to be in agreement with our Spitzer data.\\
\\
\textbf{\cite{jorgensen08}:} Building upon the \cite{lada03} definition of an embedded cluster, J{\o}rgensen et al. define a cluster as being ``loose'', which is the same as an embedded cluster, and a ``tight'' cluster. A tight cluster requires a stellar-mass volume density of $\geq$ 25 $ M_{\sun}$ $\textrm{pc}^{-3}$ and $>$ 35 YSOs, which implies that 62\% of the YSOs from our data are contained in such clusters. This finding is close to 54\% as found in \cite{evans09}.\\
\\
\textbf{\cite{gutermuth09}:} This method employs the minimal spanning tree (MST) algorithm to define {\it cluster cores} by isolating the densest parts of larger scale over-densities. The MST is a network of lines that connects a set of points, has no closed loops, and the set of edges add up to the shortest total length possible between all points. After determining a cutoff length for the MST collection, YSOs can be separated into two populations: clustered and distributed. The authors found that the clusters from this analysis range between 0.64 and 78 YSOs $\textrm{pc}^{-2}$ with a median of 60 YSOs $\textrm{pc}^{-2}$. Roughly 43\% of the YSOs are found in a median core clusters.\\
\\
\textbf{Megeath et al. in prep.:} A cluster is a set of contiguous objects which have nearest neighbour densities $\geq$ 10 YSO $\textrm{pc}^{-2}$. The 10 YSOs $\textrm{pc}^{-2}$ is similar to the cluster definition given in \cite{allen07} and motivated by a comparison of the Orion (Megeath et al. in prep.) and the Taurus molecular clouds. The Taurus and other similar dark clouds, i.e. Chameleon and Lupus, have most of their objects at densities below 10 $\textrm{pc}^{-2}$, while Orion and other clouds with clusters have 70-80\% above this threshold. Applying the 10~YSOs $\textrm{pc}^{-2}$ definition to our dataset results in $73$\% of YSOs being in clusters.\\
\\
In Fig.~\ref{fig:4box}c we show five vertical grey lines that refer to the defined densities required for a collection of YSOs to be considered ``clustered'' (Lada \& Lada~2003; Megeath et al. in prep.; J{\o}rgensen et al.~2008; Carpenter~2000, Gutermuth et al.~2009). The vertical lines fall on the same featureless slope and do not correspond to any preferred density. The black vertical line, which corresponds to dense clusters (as defined in \citealt{gutermuth05}), shows that $<26$\% of YSOs are formed in environments where they (along with their disks and planets) are likely to interact with their neighbours.

\section{Discussion and Conclusions} \label{sec:conclusions}

We have compared our global surface density distribution with previously reported definitions of clusters (discussed in \S~\ref{sec:identification}), and find that the fraction of stars in the solar neighbourhood forming in clusters is crucially dependent on the adopted definitions (ranging from $\sim40$ to $90$\%). \citet{lada03} used a physically motivated definition of clusters, and their adopted low surface density of $\sim 3$ YSO pc$^{-2}$ encompasses nearly all star formation in the solar neighbourhood. However, only a small fraction ($< 26$\%) of stars form in dense clusters  where their formation and/or evolution is expected to be influenced by their surroundings.

We conclude that stars form in a broad and smooth spectrum of surface densities and do not find evidence for discrete modes of star formation in the $\Sigma$ of low mass YSOs forming in the solar neighbourhood. Only a small fraction of YSOs form in dense clusters where nearby YSO members affect its disk/planets evolution. 
The observed lognormal surface density distribution is consistent with predictions of hierarchically structured star-formation, where the structure comes from the MC hierarchical structure \citep{elmegreen02,elmegreen08}. By hierarchical structure we mean a smoothly varying non-uniform distribution of densities, where denser subareas are nested within larger, less dense areas \citep{scalo85,elmegreen06,bastian07}. Star forming environments provide the initial conditions from which star clusters may eventually form, albeit rarely. Since the probability density function of molecular gas varies with environment, as does the tidal field experienced by the SF region, it is likely that the fraction of YSOs ending up in bound star clusters varies with environment \citep{elmegreen08} and the observed $\Sigma_{\rm YSO}$ is not universal. Hence, in a future study we will extend this work out to 2~kpc, which includes high-mass star-forming regions and more extreme environments that may show different results than what we see for the solar neighbourhood.

\section*{Acknowledgments}
We would like to thank Mark Gieles, Michael Meyer, August Muench, Dawn Peterson, Thomas Robitaille and Scott Schnee for their thoughtful feedback on the research. EB gratefully thanks the Royal Astronomical Society for their generous travel support. NB is supported by an STFC Advanced Fellowship. TPR would like to thank Science Foundation Ireland for support under grant 07/RFP/PHYF790. Support for this work, part of the Spitzer Legacy Science Program, was provided by NASA through contract 1224608 and 1288664 issued by the Jet Propulsion Laboratory, California Institute of Technology, under NASA contract 1407. NJE acknowledges support by NSF Grant AST-0607793.

\bibliography{library_mod}

\end{document}